\documentclass{sig-alternate}
\usepackage{graphicx,subfigure}
\usepackage{algorithm}
\usepackage{algorithmic}
\usepackage{flexisym}
\usepackage{dblfloatfix}
\usepackage{fixltx2e}

\DeclareMathOperator*{\mean}{mean}
\begin{document}

\title{Habits vs Environment: What Really Causes Asthma?}

\permission{Permission to make digital or hard copies of all or part of this work for personal or classroom use is granted without fee provided that copies are not made or distributed for profit or commercial advantage and that copies bear this notice and the full citation on the first page. Copyrights for components of this work owned by others than ACM must be honored. Abstracting with credit is permitted. To copy otherwise, or republish, to post on servers or to redistribute to lists, requires prior specific permission and/or a fee. Request permissions from Permissions@acm.org.}
\copyrtyr{2015}
\conferenceinfo{WebSci '15}{June 28 - July 01, 2015, Oxford, United Kingdom}
\global\copyrightetc{\copyright \the\copyrtyr\ ACM \the\acmcopyr}
\crdata{978-1-4503-3672-7/15/06\$15.00.\\   
DOI:  http://dx.doi.org/10.1145/2786451.2786481
}
\numberofauthors{3} 
%
\author{
Mengfan Tang\\
       \affaddr{Department of Computer Science}\\
       \affaddr{University of California, Irvine}\\
       \email{mengfant@uci.edu}
\alignauthor
Pranav Agrawal\\
       \affaddr{Department of Computer Science}\\
        \affaddr{University of California, Irvine}\\
       \email{pranavda@uci.edu}
\alignauthor Ramesh Jain\\
\affaddr{Department of Computer Science}\\  
       \affaddr{University of California, Irvine}\\
       \email{jain@ics.uci.edu}
}

\maketitle

 \begin{abstract}

Despite considerable number of studies on risk factors for asthma onset, very little is known about their relative importance. To have a full picture of these factors, both categories, personal and environmental data, have to be taken into account simultaneously, which is missing in previous studies. We propose a framework to rank the risk factors from heterogeneous data sources of the two categories. Established on top of EventShop and Personal EventShop, this framework extracts about 400 features, and analyzes them by employing a gradient boosting tree. The features come from sources including personal profile and life-event data, and environmental data on air pollution, weather and PM2.5 emission sources. The top ranked risk factors derived from our framework agree well with the general medical consensus. Thus, our framework is a reliable approach, and the discovered rankings of relative importance of risk factors can provide insights for the prevention of asthma.

\end{abstract}

, {Human Factors}
\section{Introduction}
Although asthma is a potentially life threatening lung disease and has been studied for a long time, asthma causes are still unclear. With increasing amount of data, such as Electronic Health Records, social media, environmental sensory data, smart-phone and wearable sensor data, a data driven solution holds the promise for helping solve this problem \cite{ram_predicting_2015}. Thus, lots of studies have been performed using various forms of data.  \cite{lee_novel_2011} predicted asthma attacks by considering bio-signals of patients and environmental data. \cite{kanani_sadat_investigating_2014} studied the association between spatial distribution of allergy prevalence and air pollutants such as PM2.5, as well as living distance from point of interests such as parks and roads. \cite{galant_predictive_2004} detected asthma risk from personal profiles. \cite{garg_multimodal_2010} designed tools for discovering dynamic changes in body sensor network data streams of asthma patients.  \cite{arif_occupational_2009} studied occupation as a factor in asthma. \cite{ho_air_2007} researched associated risk factors of air pollution and weather, \cite{hong_lifestyle_1994} revealed the impact of lifestyle and behavior on asthma vulnerability.

However, these studies only focus on either environmental factors or personal factors without a comprehensive study covering both. Asthma patients vary in sensitivity to different environmental and personal factors, and their interactions. A data driven solution can convert personal data and environmental data into information and insights for discovering comprehensive risk factors, as well as aid in understanding of asthma.

To identify these risk factors within integrated diverse data sources, new approaches are required. We propose a framework on top of Personal EventShop \cite{jalali_building_2013} and EventShop \cite{gao_eventshop:_2012}, to solve this problem because of their capability of integrating and analyzing heterogeneous data sources. EventShop is a generic infrastructure for application developers to analyze varied spatio-temporal data streams. Personal EventShop is a unified framework for aggregating personal data streams to analyze life events and personal situations. Details of this framework are described in the next section.

\begin{figure*}[t]
\centering
\label{fig:pollution}
\subfigure{\label{pm25:agri}\includegraphics[width=0.32\linewidth]{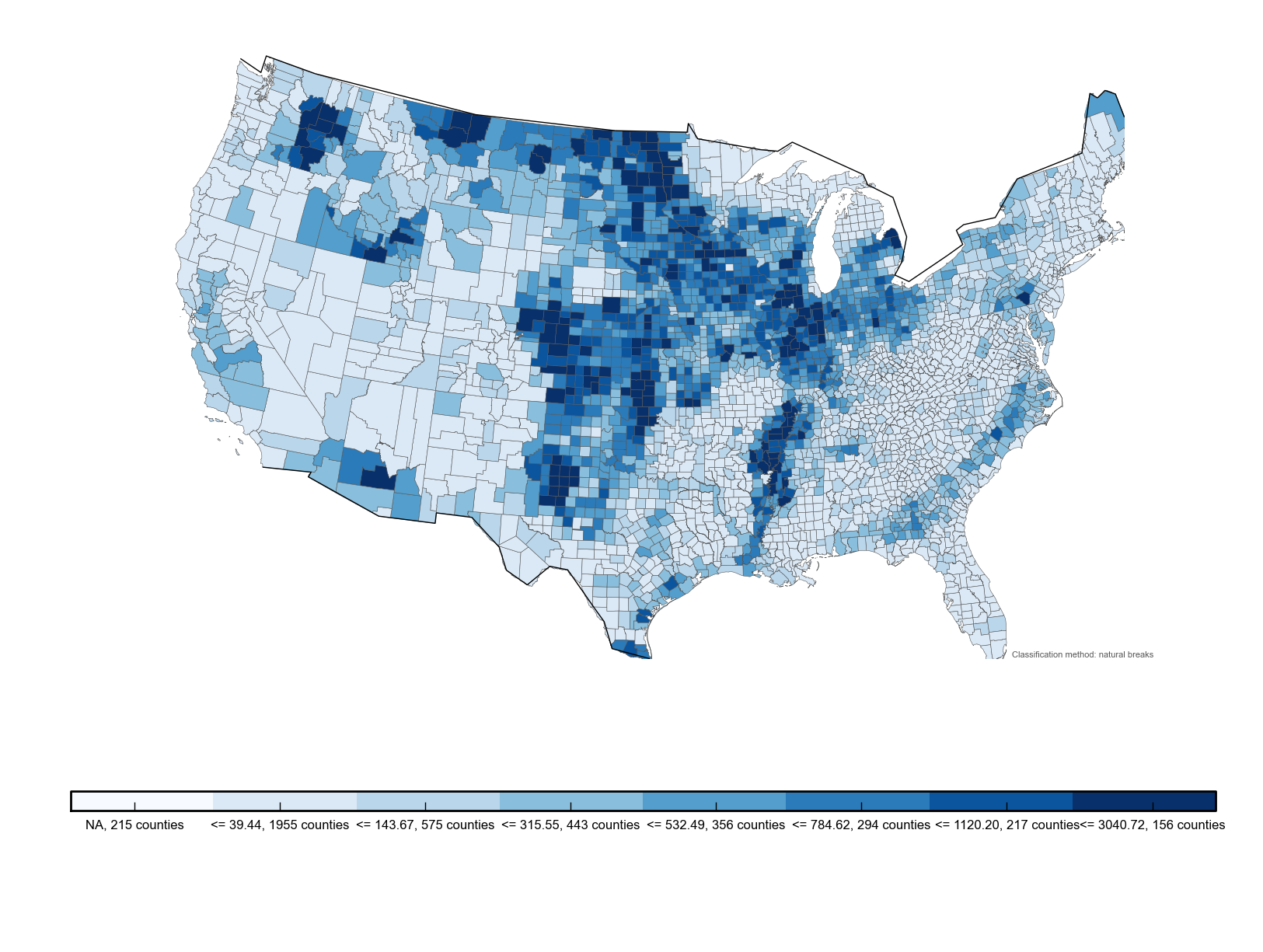}}
\subfigure{\label{pm25:dust}\includegraphics[width=0.32\linewidth]{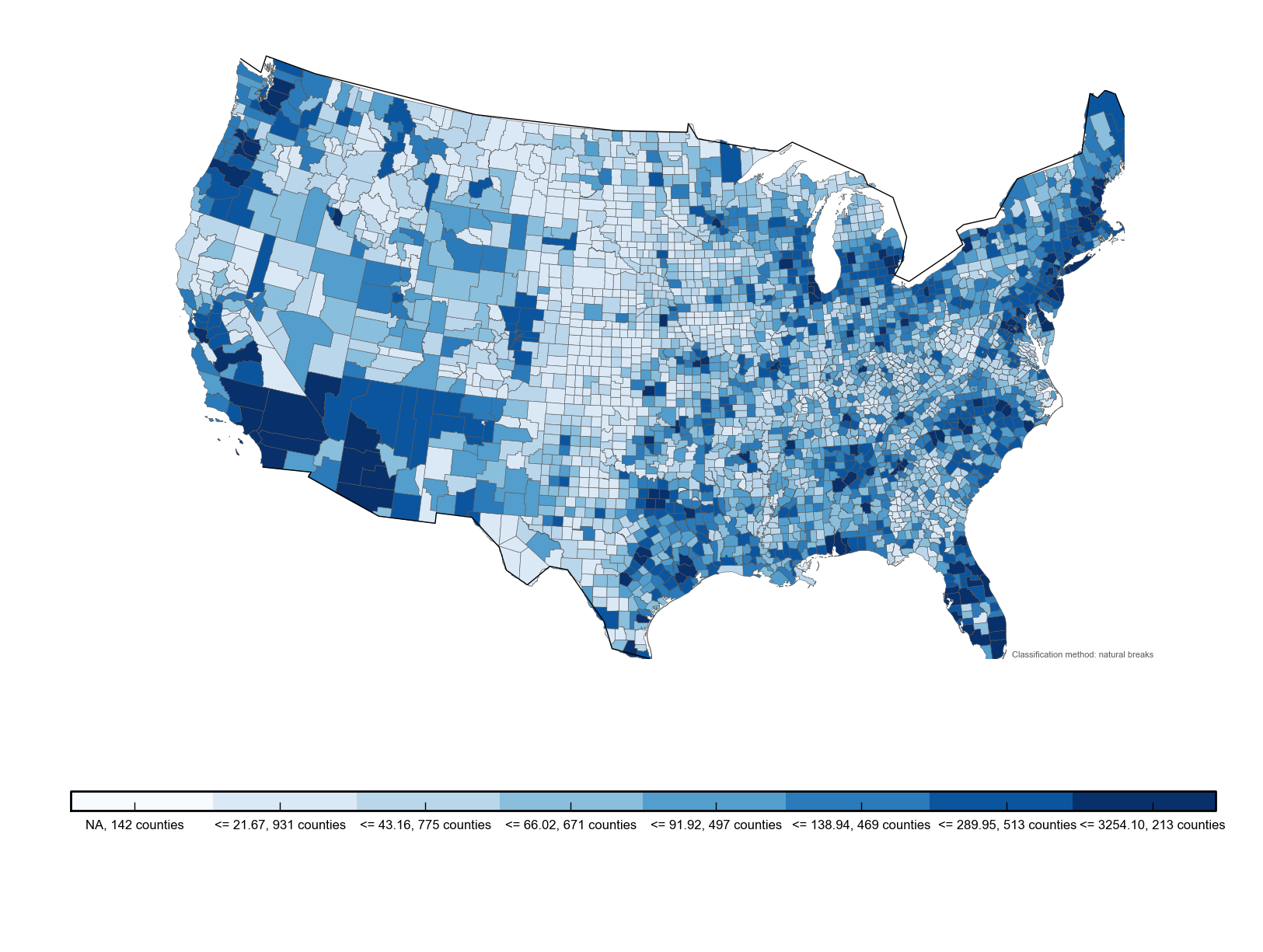}}
\subfigure{\label{pm25:mine}\includegraphics[width=0.32\linewidth]{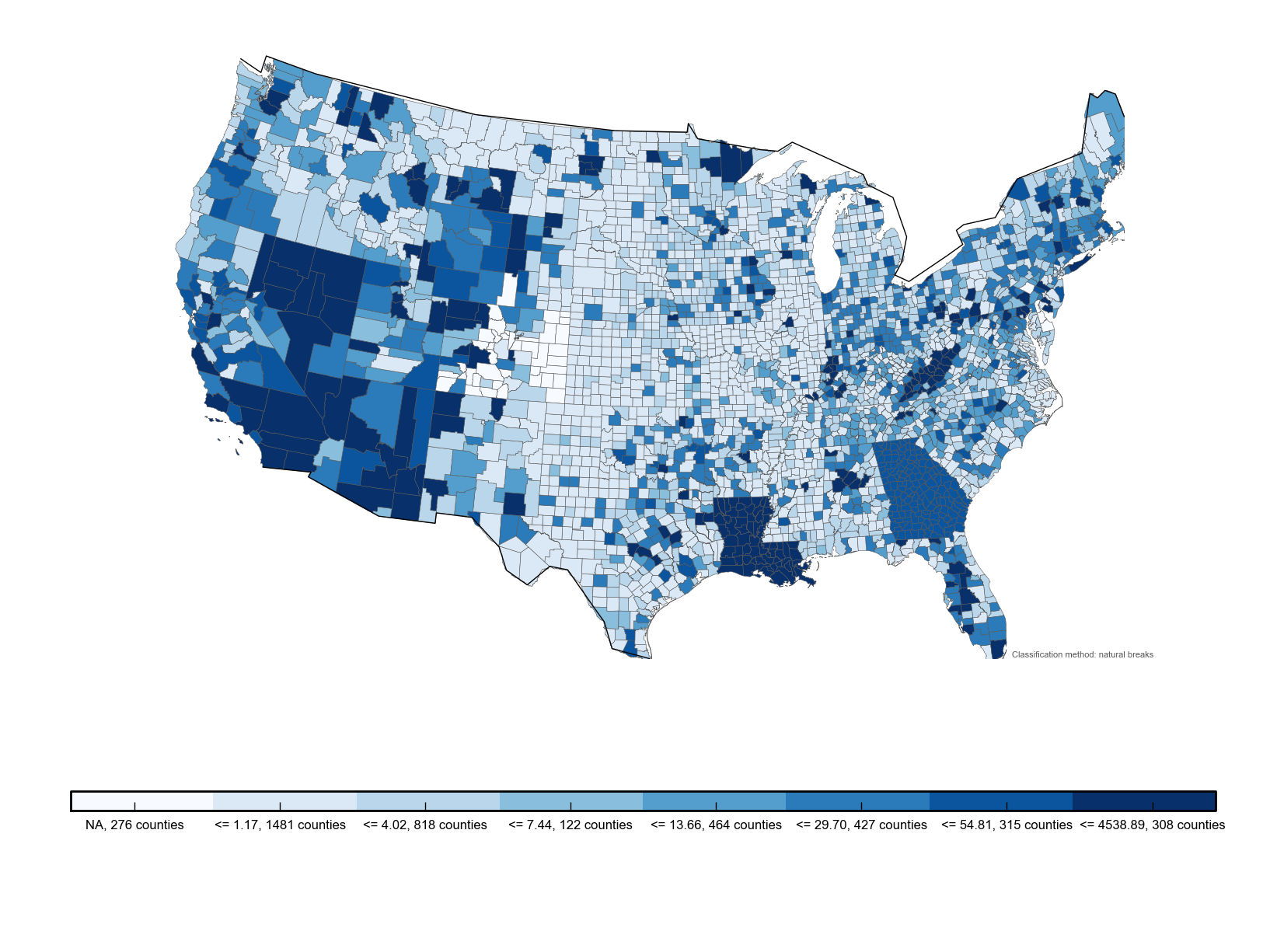}}
\subfigure{\label{pm25:max}\includegraphics[width=0.32\linewidth]{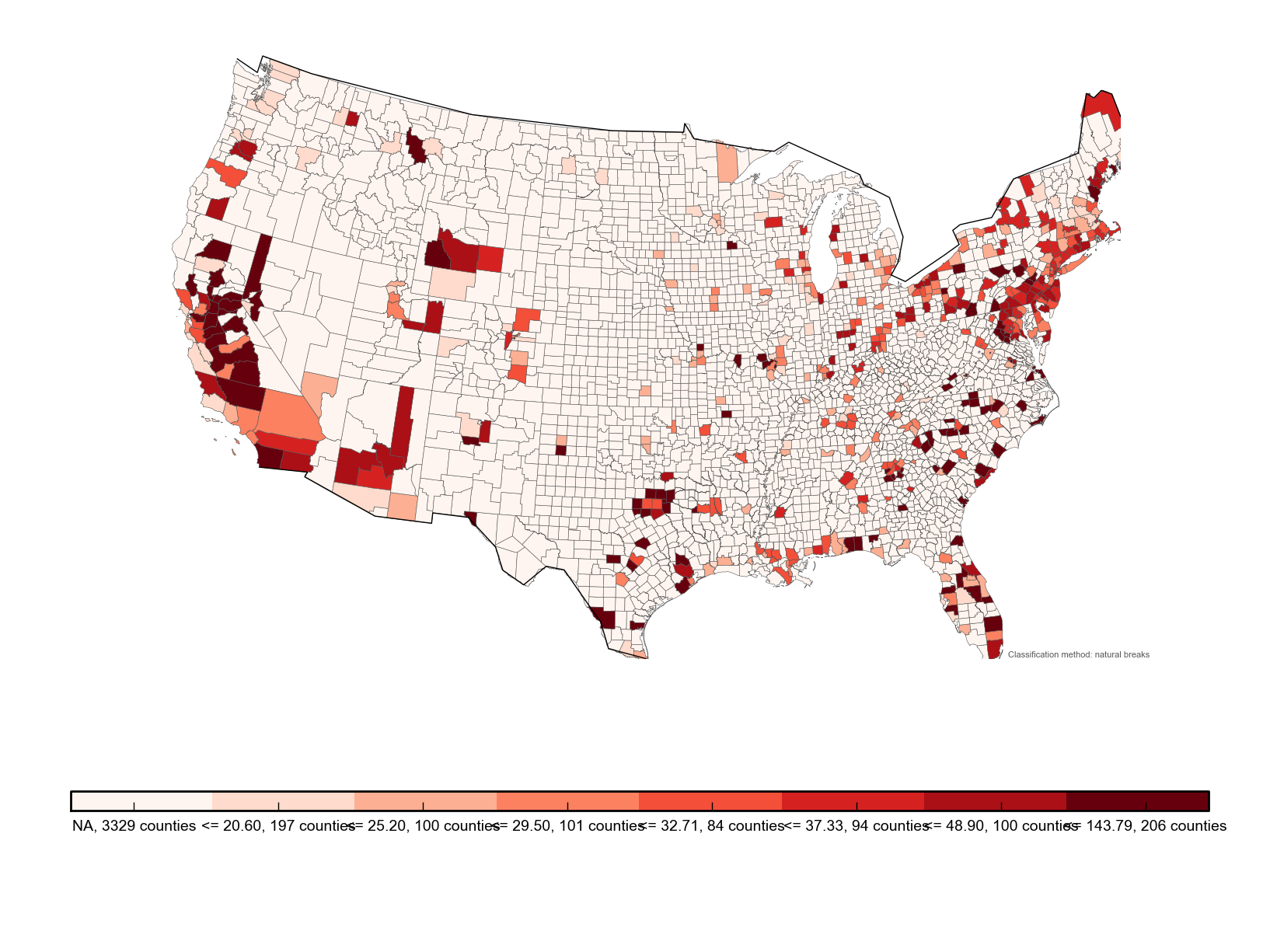}}
\subfigure{\label{pm25:mean}\includegraphics[width=0.32\linewidth]{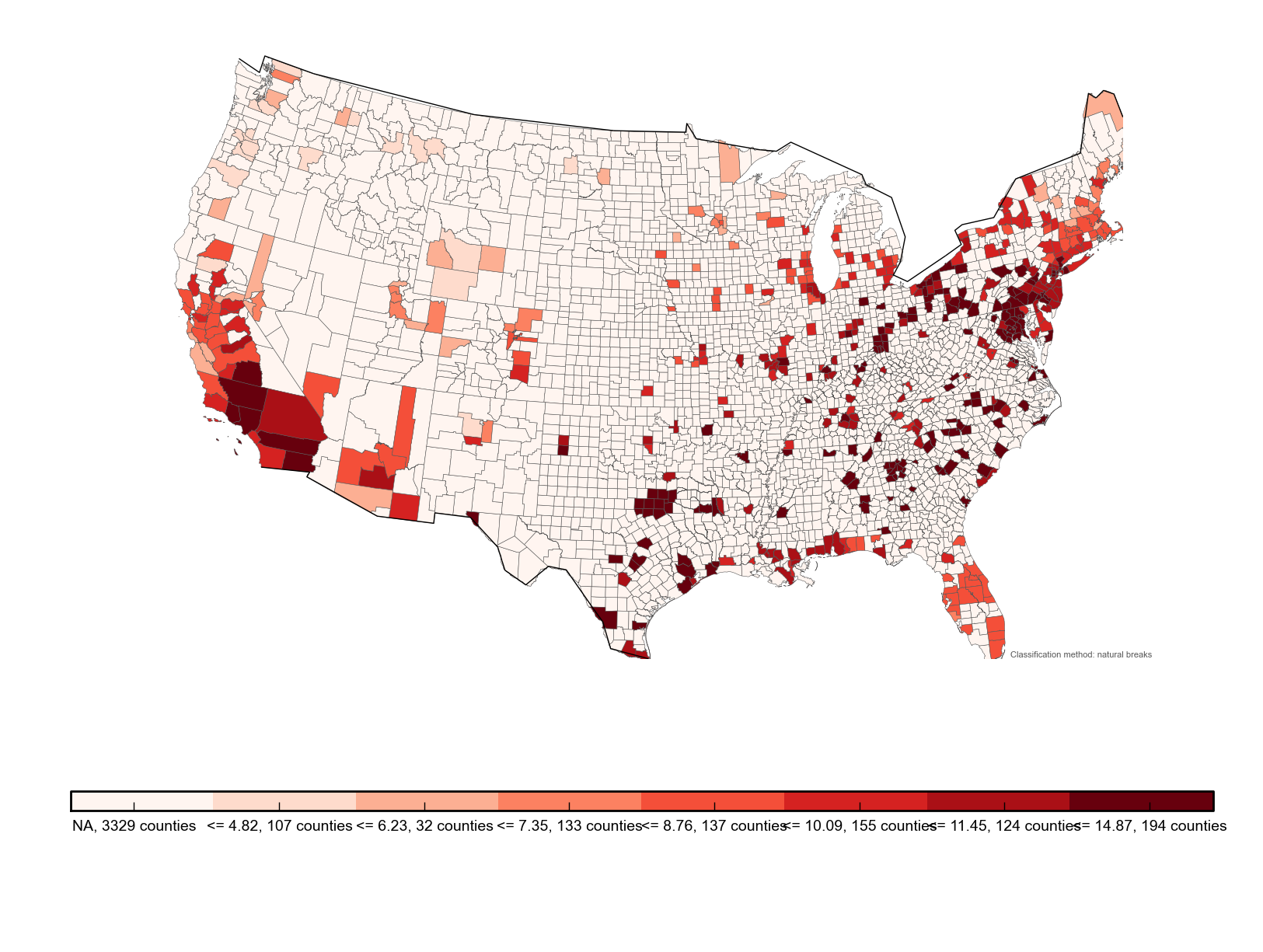}}
\subfigure{\label{pm25:min}\includegraphics[width=0.32\linewidth]{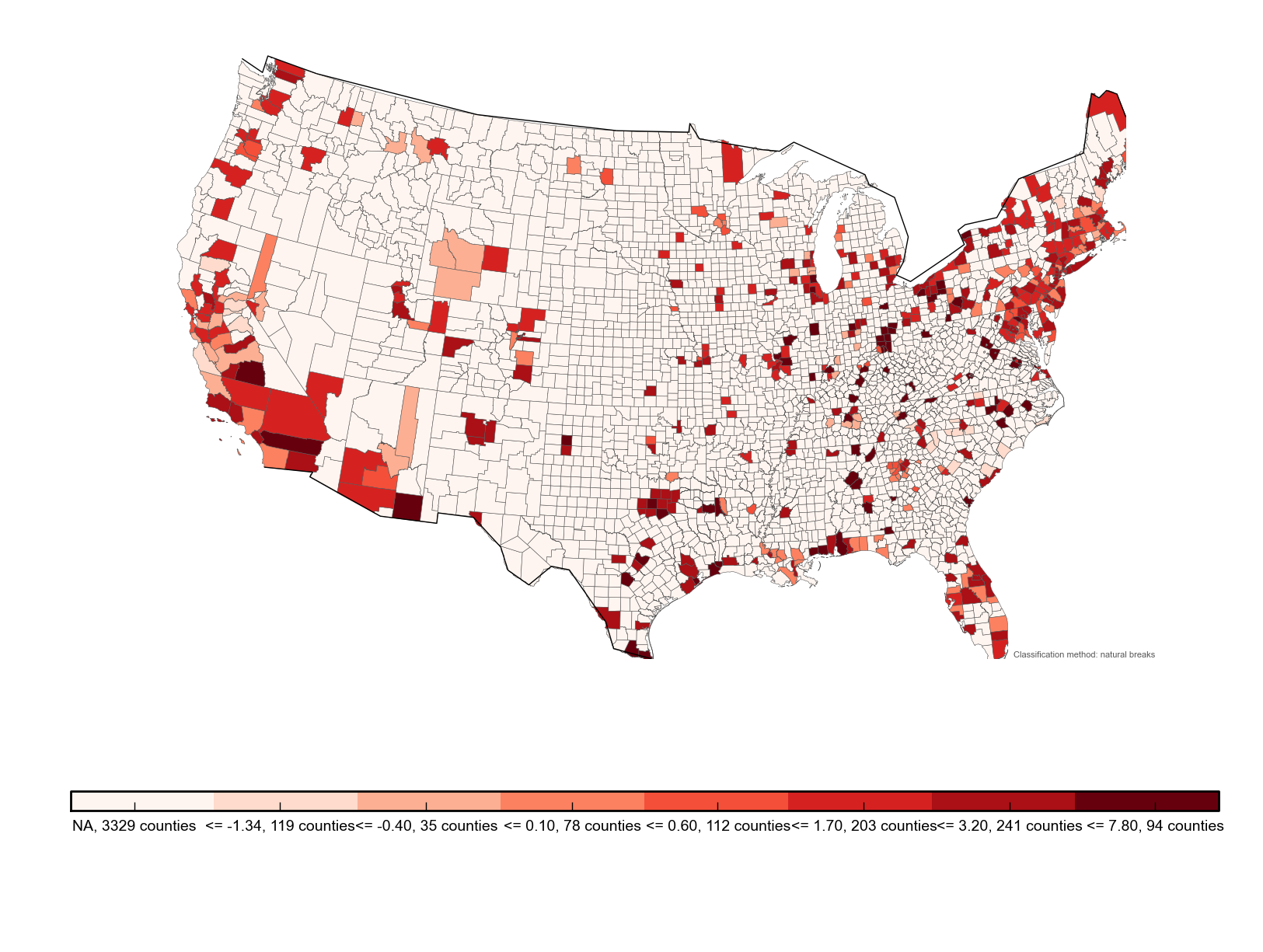}}
\caption{\ref{pm25:agri}-\ref{pm25:mine} Maps showing expected county level emissions from agriculture, road dust, and mining respectively \ref{pm25:max}-\ref{pm25:min} Maps showing average maximum, mean, and minimum PM2.5 concentration values for the month of June, for counties with data in CHAD}
\end{figure*}

\section{Framework}

The proposed framework is based on Personal EventShop and EventShop. The three main parts of the framework are briefly discussed.

\subsubsection*{Data ingestion}
Personal EventShop receives data from wearable sensors, mobile apps like calendar, smart-phone sensors like accelerometers and GPS. It correlates these data streams, determines user's activity levels, life events such as exercising, sleeping, and working. This chronicle of life events, called `Personicle' is stored in a database with attributes of user-id, location, time and life events. It also collects data from global sources like EventShop, which can provide aggregated environmental data from different sensors. In the context of this paper, EventShop can provide interpolated PM2.5 values inferred from sources like air pollution monitoring sites, satellite imagery and traffic \cite{tang_geospatial_2015}. All of this data is collected and sent for aggregation and matching.

\subsubsection*{Data aggregation}

Ingested data has a wide range of granularities in both space and time. For example, in the context of this paper, emission inventory data was reported on county level once every three years, activity data was reported over a period of several years, and varying levels of spatial granularity, and pollution data was collected at station locations, and was reported as daily values. Personal EventShop takes in all this data and matches it to a common spatio-temporal scale. This data is sent to the Analysis Engine.

\subsubsection*{Analysis Engine}

This part uses aggregated data, and performs desired analyses on it. For example, these analyses can be, co-occurrence pattern detection, personal situation recognition or life event detection. In this case, the aggregated data from the sources described in section \ref{sec:data} are transformed to features as described in section \ref{sec:features}. The features with the highest discriminatory power are detected using the model described in section \ref{sec:model} and then used to rank the asthma risk factors.

\section{Data}
\label{sec:data}
Our asthma risk analysis approach requires various personal and environmental features at the same granularity. However, the data is at different resolutions, such as point data and region data. Data sources are described in this section, and the methodology to obtain the features from the data is discussed in the section below. 

\subsubsection*{Consolidated Human Activity Data}
Consolidated Human Activity Database Master (CHAD) \cite{t_national_2000} is a collection of profile and activity information from 22 studies, and contains data for over 54,000 days combined from over 700 counties across the United states. Each personal profile contains information like zip-code, county, age, gender, race, history of asthma, cardiovascular illness, whether the person is a smoker, or lives with a smoker, employment, education level, and income. Every person has a set of diary entries. Each diary entry is an activity record containing activity (like walking, exercising, sleeping, leisure), location (home, office, traveling), start and end times. Every diary entry also contains flags showing whether the subject was breathing heavily and whether the subject was smoking during each activity.

\subsubsection*{Emission Factors}
The National Emission Inventory data was collected in 2011, by the EPA. This is an estimate of PM2.5 contributions of all the sources of air pollution within the United States, in $metric\ tonnes/year$. Sources are categorized into classes like Agricultural, Industrial, Dust, Fuel, and Mobile.

\subsubsection*{Historical Air Pollution Data}
Historical air pollution data is the actual concentrations of pollutants in the environment. This data is available as daily average values recorded at each of the individual monitoring stations across the United States. The stations also record the daily average weather conditions: temperature, pressure, wind speed, and wind direction.

In the next section, we describe the features extracted from these data sources, and how they are used.

\section{Feature and Data Analysis}
\label{sec:features}
\subsection{Personal Life Events Features: $F_{P}$}
We use the profile information provided in CHAD for people that had location information at the county level, and had answered the question whether they had asthma or not. The number of such people was 11,000, with 24,000 days of activity data. Of these, around 950 people reported having asthma, which is close to the asthma prevalence rate of 8.7\% in the United States \cite{je_national_2012}. For analysis, we use the asthmatic subjects and an equal number of non-asthmatic subjects. The non-asthmatic subjects are randomly chosen to balance the data set.

Features extracted and their justifications, as mentioned in \cite{je_national_2012} are described in Table \ref{tab:personalFeatures}. 

\begin{figure*}[t]
\label{fig:personicle}
\includegraphics[width=1\linewidth]{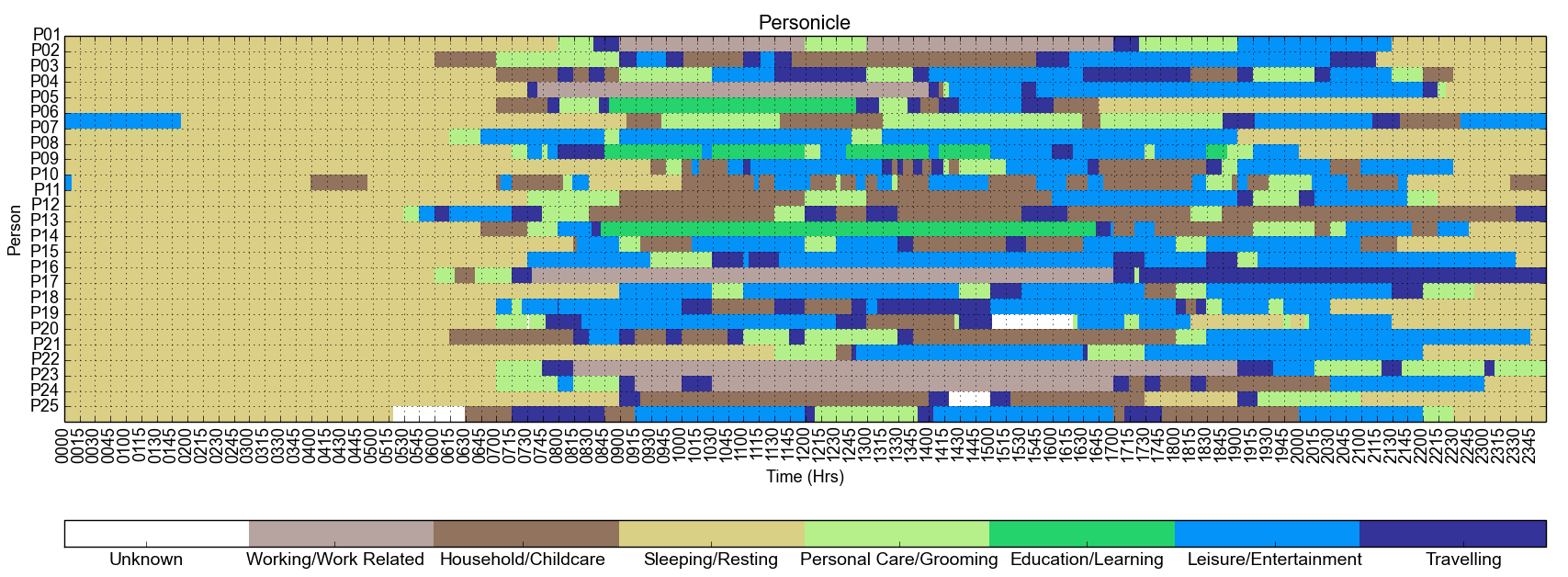}
\caption{25 randomly chosen diary entries from CHAD. Each coloured row on the y-axis represents a day's activities for one person. The color marks the activities, and the x-axis shows time in 15 minute increments}
\end{figure*}

\begin{table}[h]
\centering
\caption{Extracted Personal Profile Features}
\begin{tabular}{p{0.42\linewidth}p{0.57\linewidth}} \hline\hline
\textbf{Feature}&\textbf{Comment}\\ \hline\hline
\textbf{Age} & Children are more susceptible than adults \\ \hline
\textbf{Gender} & Females are more susceptible than males\\ \hline
\textbf{Active \& Passive Smoking} & Smokers are more susceptible \\ \hline
\textbf{Occupation, Income, \& Education} & Lower income and lower education correlate with asthma\\ \hline
\textbf{Hours of work \& Employment Status} & Measure of stress\\ \hline
\textbf{Gas stove ownership, heating \& fuel type} & Exposure to smoke\\ \hline
\hline
\end{tabular}
\label{tab:personalFeatures}
\end{table}

Activity data is in the form of a set $D$ of daily diary entry tuples. Each tuple $x$ for the day $j$ contains, the activity's start time, duration $d_{x,j}$, activity code $a_{x,j}$, location code $l_{x,j}$, flags for whether the person was smoking ($s_{x,j}$), whether the person was breathing heavily ($hb_{x,j}$) during the activity . First, locations are classified into a set $L$ of five non-exclusive categories: 
$$L=\{work,\ travel,\ home,\ indoor,\ outdoor\}$$
Activities are classified into a set $A$ of six non-exclusive categories:
$$A= \{sleep,\ work,\ exercise,\ walking,\ cycling,\ leisure\}$$ 

For every person $p_i$ for day $j$, in the set of people $P$, let the set of corresponding diary entries be $D_{p_i}^j$. For every person $p_i$, and every location category $L_k$, the average time spent daily at locations in that category $t_{L_k}^i$ is computed for diary entries $x$ in $D_{p_i}^j$ over all $j$:
$$t_{L_k}^i= \mean_j(\sum_x d_{x,j}) \qquad\{\forall j,x : x \in D_{p_i}^j\ and\ l_{x,j} \in L_k \}$$
Similarly, for activities, the average time spent daily by $p_i$ performing activities in each activity category $A_k$, denoted by $t_{A_k}^i$, is computed. Also, the average amount of time in the entries where $p_i$ reports heavy breathing, is computed as $t_{hb}^i$:
$$t_{hb}^i=\mean_j(\sum_x d_{x,j}) \qquad\{\forall j,x : x \in D_{p_i}^j\ and\ hb_{x,j} = 1\}$$
The average amount of time in the entries where $p_i$ reports smoking, is also calculated as $t_{s}^i$, in the same way.

The average count of the number of times $p_i$ reports heavy breathing, is recorded as ($n_{hb}^i$):
$$n_{hb}^i=\mean_j(\sum_x hb_{x,j}) \qquad\{\forall j,x : x \in D_{p_i}^j\ and\ hb_{x,j} = 1\}$$
Likewise, the average count of activities per day, where the smoking flag was 1, is calculated as $n_{s}^i$.
Figure \ref{fig:personicle} shows twenty five randomly chosen daily diaries from the data.

\subsection{Emission Factors Features: $F_{E}$}

PM2.5 emission sources from the national emission inventory were used at county level. Different kinds of PM2.5 emissions have different health effects on people. The emission amount by each type of pollution source, like mobile, industrial, dust, fires, and fuel, was used as a feature. The various emission factors are listed in Table \ref{tab:emissionFactors}.

\begin{table}
\centering
\caption{Emission Factors}
\begin{tabular}{c p{0.62\linewidth}}
  \hline\hline
  \textbf{Category} & \textbf{Emission Factor} \\
  \hline \hline
  \textbf{Mobile} & Aircraft,
   Marine Vessels,
   Locomotives,
   Equipment,
   Heavy Duty Vehicles,
   Light Duty Vehicles\\
  \hline
  \textbf{Industrial} & Agricultural, 
   Mining,
   Oil \& Gas Production,
   Storage \& Transportation, 
   Other\\
  \hline
  \textbf{Dust} & Construction, 
   Paved Road Dust, 
   Unpaved Road Dust\\ 
  \hline
  \textbf{Fires} & Agricultural Field Burning, 
   Prescribed Fires, 
   Wildfires \\
  \hline
  \textbf{Fuel} & Biomass,
   Coal,
   Natural Gas,
   Oil, 
   Residential Wood, 
   Other \\
  \hline 
  \textbf{Miscellaneous} & Waste Disposal,
   Agriculture,
   Commercial Cooking \\
  \hline \hline
\end{tabular}
\label{tab:emissionFactors}
\end{table}

\subsection{Air Pollution Features: $F_{A}$}
Historical air pollution data records are in the form of daily average pollutant concentration values at the measuring station level. Since all of our analysis is at county level, we interpolate this point data to region data. This can be done using a lot of techniques, ranging from pure chemical transportation models, to enhanced data based techniques \cite{tang_geospatial_2015}. For this experiment, linear interpolation was used.

The data used is between years 2001 and 2014.
The pollutants considered are: PM2.5, Ozone, Carbon Monoxide, Sulphur Dioxide, and Nitrogen Dioxide. The weather factors considered are: temperature, pressure, and wind speed.

For each pollutant and weather factor $f$, the following features are extracted:
$$f_{max}^m=\mean_y(\max_d(v_{d,f}^{m,y}))$$
$$f_{mean}^m=\mean_y(\mean_d(v_{d,f}^{m,y}))$$
$$f_{min}^m=\mean_y(\min_d(v_{d,f}^{m,y})),$$
where, $f \in \{PM2.5,\ SO_2,\ NO_2,\ O_3,\ CO,\ temperature,\\ pressure,\ wind speed\}$, $y$ is the year between 2001 and 2014, $m$ is the month and $d$ is the day of the month.

\section{Models}
\label{sec:model}
Consider a feature vector $x\in \mathbb{R}^d$ in the $d$-dimensional feature space, $y^{+}$ is the label for subjects with asthma, $y^{-}$ is the label for subjects without asthma. Our feature ranking model is based on gradient boosting regression tree \cite{friedman_greedy_2001}. 
Risk factors are ranked by analyzing importance of features from this model. Gradient boosting tree framework generates an ensemble of weak regression tree models and combines all the weak learners to produce a strong classifier. The training algorithm performs gradient descent in function space to minimize a differentiable loss function. Suppose $F(x)$ is the model, we have,
\begin{equation}
F(x) = \sum_{i=1}^M \lambda_{i} f_{i}(x)
\end{equation}
where $f_{i}(x)$ is the $i^{th}$ weak learner, $\lambda_{i}$ is the weight associated with that weak learner.  

One of the advantages of using this model for ranking features in asthma study is that no normalization of data is needed, and thus it is better at handling categorical data with discrete and continuous data. 

The model hyper-parameters for the gradient boosting classifier (depth and number of trees) are selected using five-fold cross validation. The depth was varied between 1,2, and 3, and the number of trees was chosen from 50, 100, and 150. After choosing the best hyper-parameters, the learned model was then applied to the remaining subset.

A $K$ nearest neighbor classifier was used to predict whether a person is asthmatic or not, using their features. The parameter K was, again, cross validated on the training data using five folds, to select the best K.
\begin{figure}[h]
\centering
\epsfig{file= 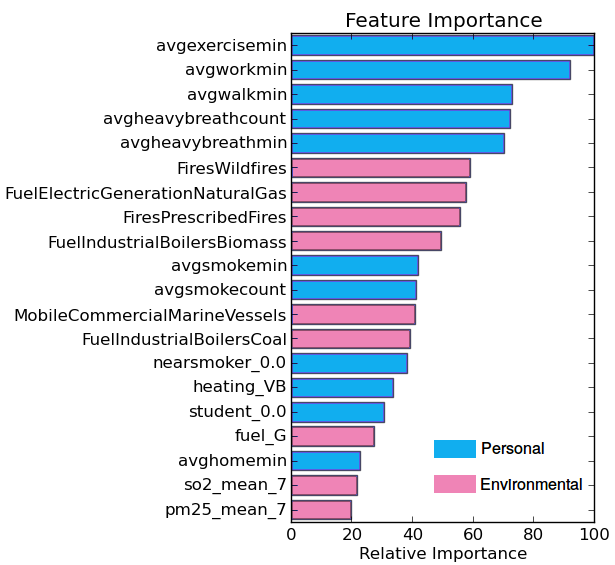,width=1\linewidth}
\label{fig:feateff}
\caption{Relative importance of features as determined by a gradient boosting tree}
\end{figure}

\section{Results}

In this section we present our results based on the features extracted from the data sources, showing the effectiveness of the identified data sources.

The most significant contribution of our paper is the determination of the relative importance of factors in causing asthma. The twenty most effective features out of around 400 extracted features, and their relative importance is shown in Figure \ref{fig:feateff}. The top five features correspond to personal factors such as physical exertion and work stress. Following that, environmental features like forest fires and industrial fuel burning show up. Other features that appear in the analysis are smoking habits, passive smoking, personal exposure to smoke from domestic fuels and heating. Lastly PM2.5 and SO$_2$ concentration during summer (July) also appear on the list. These are all, independently, known to be important factors affecting asthma susceptibility. Our ordering suggests that personal factors like physical exertion and stress are strong indicators of asthma risk, followed by exposure to PM2.5 and smoke from the environment. The personal features obtained here are particularly useful because they don't need extensive diary keeping by the subjects. Most of them, like physical exercise time, work hours, time at home, and breath rate can be easily obtained from wearable sensors and smart-phones. Other personal features like, smoking habits, fuel type, and occupation can be obtained from one time questions.

The effectiveness of our features is shown by the good performance of an asthma classifier trained on them. Four K-nearest neighbors classifiers are trained on the various subsets of the data: one on just $F_P$, one on $F_P\ and\ F_A$, one on $F_P\ and\ F_E$ and one on all, $F_P,F_E\ and\ F_A$. The Area Under the Curve metric, which is a commonly used metric in clinical studies \cite{zweig_receiver-operating_1993}, is used to evaluate the performance of the classifier. The performance is shown in Table \ref{tab:feature_effectiveness}. The table clearly shows, that asthma prediction based on both environmental features and personal features, significantly outperforms using any subset of these features. It may be noted that the recall of the system improves significantly, on moving from $F_P +F_E$ to $F_P+F_E+F_A$, at the cost of a small drop in precision.
\begin{table}[h]
\centering
\caption{Classifier performance against feature set}
\label{tab:feature_effectiveness}
\begin{tabular}{|c|c|c|c|} \hline
\textbf{Features}&\textbf{Precision}&\textbf{Recall}&\textbf{AUC}\\ \hline
$F_{P}$ & 0.818&0.789&0.807\\ \hline
$F_{P}$+$F_{A}$&0.846&0.864&0.853\\\hline
$F_{P}$+$F_{E}$& \textbf{0.902}&0.807&0.860\\\hline
$F_{P}$+$F_{E}$+$F_{A}$&0.898&\textbf{0.927}&\textbf{0.911}\\\hline
\end{tabular}
\end{table}
\section{Conclusions}
There is no clear understanding of the causes of asthma and no definite cure, making it draw increasing attention in health care studies. Existing studies often are performed on either personal data or environmental data. We proposed a framework on top of EventShop and Personal EventShop, integrating environmental data and personal profile and activity data, and extract personal and environmental features that determine a person's susceptibility to asthma. A ranking of these features is given based on their potential in causing asthma. The features derived from our analysis agree with the current general consensus about factors affecting asthma, in the medical community. 

One of the directions for future work is to extend the framework to a personalized system for individual patients by taking in their daily activity data, through Internet connected devices, and environmental data about spatio-temporal factors causing asthma attacks. This can be, then used to determine personalized high risk zones for the individual, and personalized risk factor profiles.
\section{Acknowledgments}
We thank Xikui Wang for helping extract air pollution features.
\bibliographystyle{abbrv}
\bibliography{asthma2}

\end{document}